\pdfoutput=1
\documentclass[10pt,journal,compsoc]{IEEEtran}

\usepackage{cite}
\usepackage{amsmath,amssymb,amsfonts}
\usepackage{algorithmic}
\usepackage{array}
\usepackage{textcomp}
\usepackage{stfloats}
\usepackage{url}
\usepackage{verbatim}
\usepackage{enumitem}
\usepackage[table]{xcolor}
\usepackage{booktabs}
\usepackage{tabularx}
\usepackage{bussproofs}
\usepackage{graphicx}
\usepackage{xurl}
\usepackage{mdwlist}
\usepackage{multirow}
\usepackage{subfig}
\usepackage{float}
\usepackage{adjustbox}
\usepackage{url}
\usepackage{color}
\usepackage{alltt}
\usepackage{xspace}
\usepackage{tikz}

\usepackage[ruled,linesnumbered]{algorithm2e}

\SetCommentSty{mycommfont}
\usepackage{tcolorbox}
\tcbuselibrary{listings, skins, breakable}

\newtcolorbox{promptbox}[1][]{
  colback=gray!5,
  colframe=gray!60,
  fonttitle=\bfseries\footnotesize,
  title={#1},
  breakable,
  left=4pt, right=4pt, top=2pt, bottom=2pt,
  boxsep=1pt,
  boxrule=0.5pt,
  sharp corners,
}
\SetKwInput{KwInput}{Input}
\SetKwInput{KwOutput}{Output}
\SetKwProg{Fn}{Function}{}{end}

\usepackage[bookmarks=false]{hyperref} 
\hypersetup{
    citebordercolor={0 1 0},
    citecolor=blue,
    colorlinks=false,
    filebordercolor={0 .5 .5},
    filecolor=green,
    linkbordercolor={1 0 0},
    linkcolor=blue,
    menubordercolor={1 0 0},
    pageanchor=true,
    pagebackref=true,
    pdfborder={0 0 0},
    pdfpagelabels=true,
    pdftex,
    plainpages=false,
    urlbordercolor={0 1 1},
    urlcolor=blue,
}

\usepackage{syntax}
\usepackage{multicol}
\usepackage{wrapfig}
\usepackage{float}

\newcommand{\Comment}[1]{}

\newtheorem{definition}{Definition}

\newcommand{\SmallSpace}{\vspace*{-1.5ex}}

\newcommand{\eat}[1]{}

\newcommand{\distance}{3pt}
\usepackage{listings}
\definecolor{mygreen}{rgb}{0,0.6,0}
\definecolor{mygray}{rgb}{0.5,0.5,0.5}

\newcommand{\incode}[1]{\lstinline{#1}}

\newcommand{\sys}{\textsc{AdWise}\xspace}
\newcommand{\uinav}{ad-oriented UI exploration\xspace}

\def\BibTeX{{\rm B\kern-.05em{\sc i\kern-.025em b}\kern-.08em
    T\kern-.1667em\lower.7ex\hbox{E}\kern-.125emX}}

\begin{document}

\title{From UI to Code: Mobile Ad Regulation via LLM Unified Static--Dynamic Analysis}

\author{Shang~Ma, Yanfang~Ye, and Xusheng~Xiao%
\thanks{Shang Ma and Yanfang Ye are with University of Notre Dame.
E-mail: \{sma5,yye7\}@nd.edu.}%
\thanks{Xusheng Xiao is with Arizona State University.
E-mail: xusheng.xiao@asu.edu.}%
\thanks{Yanfang Ye and Xusheng Xiao are the corresponding authors.}}

\markboth{IEEE Transactions on Software Engineering}%
{Ma \MakeLowercase{\textit{et al.}}: From UI to Code}

\maketitle

\begin{abstract}
Mobile advertisements (ads) fuel the app economy, yet poorly regulated ads can degrade user experience and expose users to malware.
Identifying ad regulation violations is challenging because ads are dynamically retrieved from remote servers and embedded across various user interfaces (UIs), making them difficult to locate and trigger. To address these challenges, we propose \sys, which innovatively harnesses Large Language Models (LLMs) to unify static ad analysis and dynamic UI exploration for detecting mobile ads and assisting ad regulation. 
In particular, \sys first develops novel static analysis to identify UI widgets used to place ads (referred to as \textit{ad widgets}).   
To effectively reach these widgets, \sys instructs an LLM to guide dynamic exploration using three distinct types of guidance derived from static analysis and ad placement practices: (1) \textit{WTG-based Guidance}, which utilizes a statically built window transition graph (WTG) to delineate the app's structural topology; (2) \textit{Semantic Guidance}, which reasons about app functionality to prioritize paths users are likely to visit (e.g., ``play" buttons in music apps); and (3) \textit{Structural Guidance}, which leverages Retrieval Augmented Generation (RAG) to recognize common ad-heavy UI layouts (e.g., settings menus) from a knowledge base. 
Our evaluation on $100$ benchmark apps shows that \sys outperforms the strongest baseline by $25.83$ percentage points in ad-widget detection and uncovers $34.34\%$ more violations across six violation types in three regulation categories, demonstrating its effectiveness for ad regulation. A broader in-the-wild study identified $383$ regulation violations within $2,047$ apps using Meta Ads. We have reported the findings to Google Play and the Abuse Track of the \textit{Meta Bug Bounty} program. 


\end{abstract}



\section{Introduction}
\label{sec:intro}

\begin{figure}[t]	
\includegraphics[width=\columnwidth]{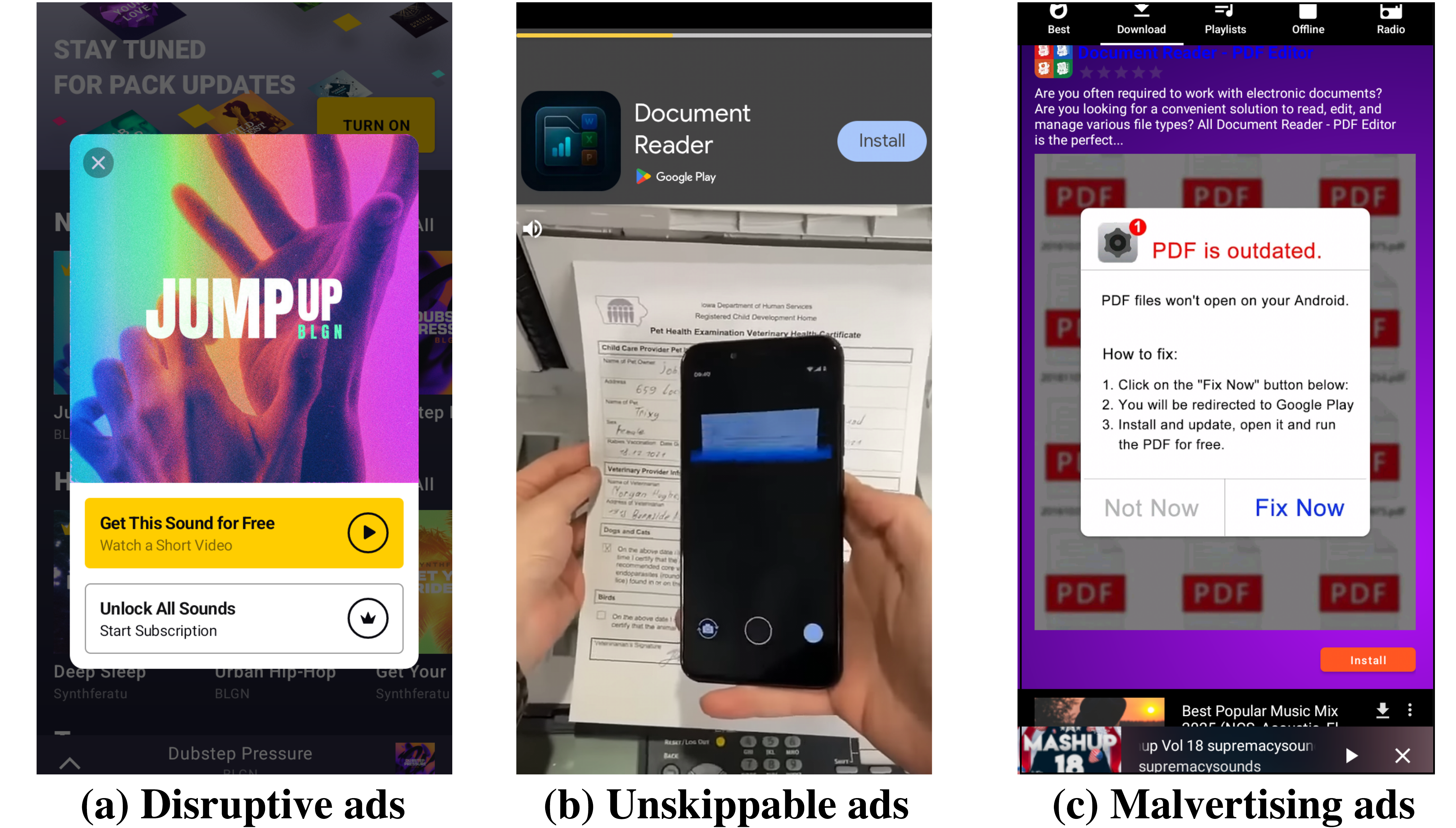}
	\caption{Examples of ad regulation violation}
	\label{fig:motivating example}
\end{figure}

Mobile advertisements, commonly known as mobile ads, are prevalent in over $57\%$ of Google Play apps~\cite{viennot2014measurement}, serving as a key source of engagement and revenue for developers~\cite{applovinsuccessstories,admobadvantages}. 
While ads help monetize apps and promote app downloads, unregulated ad practices can degrade user experiences~\cite{yan2023adhere,betteradstandards,zhao2023mobile} and compromise app ecosystem integrity~\cite{rastogi2016these,liu2020maddroid,xing2015understanding}. 
For instance, as shown in ~\autoref{fig:motivating example}, users may encounter disruptive ads that appear unexpectedly during loading, unskippable ads that trap users, or malvertising ads with deceptive content that leads to malware downloads.
Although mobile platforms and ad libraries attempt to curb these behaviours through strict policies, vetting ad compliance at scale is non-trivial,  and many apps that violate ad regulations manage to bypass review processes and reach a wide user base. 



Recognizing this problem, prior research has proposed dynamic UI exploration-based methods to detect mobile ad regulation violations~\cite{liu2020maddroid,ma2024careful,rastogi2016these,cai2023darpa,zhao2023mobile}, but these efforts rely on a content-based detection workflow that is inherently inefficient and brittle.  
 Specifically, they (1) navigate through the app using general-purpose UI exploration techniques and (2) check regulation violations by verifying each UI content against predefined rules.   
However, general-purpose explorations are optimized for broader code coverage and thus waste resources on UI exploration paths that do not lead to ads~\cite{liu2022promal,ran2023badge,ran2024guardian}.  
Second, relying solely on content ignores underlying code behaviors and fails to generalize when UI content is dynamic.  
For example, the ads in \autoref{fig:motivating example}~(b) and (c) will show different content after reloading the app, easily bypassing the content check. 


\noindent\textbf{Our Insights.}
A key to overcoming this challenge lies in \textit{steering UI exploration to pinpoint \textbf{ad widgets} (i.e., the UI components rendering ads), and rigorously examining both their code behaviors and content for signs of non-compliance}.  
Fundamentally, these widgets serve as the central mechanism that determines where ads appear, how users interact with them, and what content is delivered upon triggering.    
 To achieve this, we propose \sys, a novel approach that \textit{harnesses Large Language Models (LLMs) to unify static ad analysis and dynamic UI exploration for effective ad widget detection and mobile ad regulation}. In particular, while static analysis is capable of identifying potential ad widgets (i.e., \textit{what} to look for), it often generates infeasible paths to the UIs containing these widgets (i.e., \textit{where} to find them)~\cite{liu2022promal,gator2}. Such imprecision renders a straightforward integration of static analysis and dynamic UI exploration ineffective, as it misguides exploration along infeasible paths.

We bridge this gap by identifying two fundamental patterns in ad widget placement that current static analysis fails to capture: (1) \textit{Semantic Pattern (The ``User" Perspective)}. Ad widgets are often placed on exploration paths directly tied to the app's specific purpose and functionality, which are frequently visited by the app’s target users (e.g., popup ads appearing within a music app's playlist). (2) \textit{Structural Pattern (The ``Developer" Perspective)}. Conversely, many ad widget placements ignore specific app semantics and instead rely on generic, repetitive UI layouts common across the ecosystem (e.g., interstitial ads during loading states, or standard settings menus). Unfortunately, the varying functionalities and diverse UIs of apps make it challenging to establish consistent rules for finding ad widgets placed by these two patterns.

Our key insight is that LLMs are uniquely positioned to address this duality due to their extensive training on real-world UIs and code. They can \textit{comprehend an app's metadata from the app market} to infer user intent (addressing the Semantic Pattern) and \textit{identify similar UI layouts} from information provided by Retrieval Augmented Generation (RAG) to recognize developers' conventions (addressing the Structural Pattern). \textit{By leveraging this capability, \sys can effectively unify static ad analysis with dynamic exploration to pinpoint ad widgets.}


\noindent\textbf{Our Approach.} 
Based on the above insights, \sys is designed to have two phases:

\noindent\textit{Phase 1: Static Ad Analysis.} 
\sys employs a novel static analysis to identify different types of ad widgets based on both UI attributes and code behaviors. 
Specifically, \sys first conducts an over-approximate dataflow analysis to associate each widget with its view class, resource-id, and method invocation, and then applies inference rules along with the dataflow analysis to detect these ad widgets based on their unique characteristics of ad-showing logic (i.e., their view classes and associated method invocations).  
\sys further builds a static WTG that represents transitions among UIs (or called windows), locates the ad widgets in the WTG's UIs using their resource-ids, and computes the shortest UI exploration paths to reach these UIs.

\noindent\textit{Phase 2: Ad-Oriented UI Exploration.}  
\sys instructs LLMs to guide UI exploration to find ad widgets by digesting both static analysis results and the knowledge of placement patterns.  
Specifically, when \sys visits a UI, it first searches and triggers ad widgets in the UI based on their resource-ids, collecting the resulting ad content. 
\sys then instructs the LLM to select a UI widget (such as a button) from the UI,
and triggers the UI widget to explore more UIs.
In particular, \sys computes three types of contextual information as the exploration guidance that assists LLMs in finding paths toward ad UIs:

\begin{itemize}[noitemsep, topsep=1pt, partopsep=1pt, listparindent=\parindent, leftmargin=*]
\item \textit{WTG-based Guidance:} WTG contains UI transitions and the exploration paths. 
\sys encodes this useful information by embedding WTG into text sequences.
\item \textit{Semantic Guidance:} 
To find ad widgets adhering to the \textit{semantic pattern}, \sys analyzes an app’s metadata, such as its name, category, and description, collected from the app market to infer its functionality. 
\sys then instructs the LLM to determine which UI widgets the app’s target users are most likely to engage with. 
For instance, in audio-related apps, the generated guidance ``listening to the songs and exploring high-quality content'' enables LLM prioritize app-specific interaction, such as clicking ``Play'' or ``Next Song'' buttons, which are more likely to navigate to ad UIs. 
\item \textit{Structural Guidance:} 
To find ad widgets adhering to the \textit{structural pattern}, \sys leverages a knowledge base of standard ad UIs. Specifically, for the currently visiting UI, \sys retrieves the top $k$ most similar UIs from the knowledge base based on the similarity of the UI embeddings.
Since these retrieved UIs are known to contain ad widgets, the widgets in the current UI that structurally correspond to them have a higher likelihood of leading to ads. Thus, \sys generates summaries of the widgets from these similar UIs, helping the LLM prioritize the UI widgets most relevant to these established conventions. 

\end{itemize}

\sys iteratively explores the app by triggering UI widgets and collecting the resulting ad content, continuing this process until all ad widgets are discovered or a predefined exploration step limit is reached.
After the exploration phase, \sys applies a set of detection rules on the attributes of ad widgets to identify violations of mobile ad regulations.

This paper makes the following main contributions:
\begin{itemize}[noitemsep, topsep=1pt, partopsep=1pt, listparindent=\parindent, leftmargin=*]
    \item A novel approach named \sys that harnesses LLMs to unify static ad analysis and dynamic UI exploration for effective mobile ad regulation.  
    \item A novel static analysis technique that builds WTGs and identifies ad widgets among the windows of the WTGs.
    \item Extensive evaluation on $271$ ad widgets shows that \sys outperforms six baseline methods~\cite{monkey,ma2024careful,liu2020maddroid,cai2023darpa,lai2019goal,ran2024guardian}, improving the detection rate by at least $25.83$ percentage points and running at least $8.68\%$ faster. Meanwhile, \sys successfully identifies $133$ regulation violations, exceeding the best-performing baseline by at least $34.34\%$. These results demonstrate \sys’s effectiveness in mobile ad detection and its practical value for supporting ad regulation. 
    
\item An in-the-wild study on $2,047$ real-world apps demonstrates \sys's practical effectiveness, identifying $383$ ad regulation violations. These results are subsequently reported to Google Play and the Specialized Abuse Track of the \textit{Meta Bug Bounty} program.
\end{itemize}
\newcommand{\yes}{\scalebox{1.5}{$\bullet$}} 
\newcommand{\no}{\scalebox{1.5}{\color{gray}$\circ$}} 
\begin{table}[t]
\centering
\small
\setlength{\tabcolsep}{4pt}
\caption{Comparison of industrial requirements and academic approaches to mobile ad regulation (\yes = present, \no = absent). Five major platforms uniformly require all three categories; \sys is the only academic approach that does too.}
\label{tab:adpolicy}
\begin{tabular}{@{}l ccc@{}}
\toprule
& \textbf{Placement} & \textbf{Interaction} & \textbf{Content} \\
\midrule
Industry~\cite{google-ad-policy,apple-guideline,metaadpolicy,unity-policy,applovin-policy} & \yes & \yes & \yes \\
\midrule
FraudDroid~\cite{dong2018frauddroid} & \yes & \yes & \no  \\
MadDroid~\cite{liu2020maddroid}      & \no  & \no  & \yes \\
AdGPE~\cite{ma2024careful}           & \no  & \no  & \yes \\
AdRambler~\cite{zhao2023mobile}      & \no  & \no  & \yes \\
DARPA~\cite{cai2023darpa}            & \no  & \no  & \yes \\
\midrule
\rowcolor{gray!15}
\textbf{\sys (Ours)}                 & \textbf{\yes} & \textbf{\yes} & \textbf{\yes} \\
\bottomrule
\end{tabular}
\end{table}






\section{Background and Motivation }
 \label{sec:motivationionalstudy}

\subsection{Ad Regulation Categorization} 
Based on ad policies from Google and Apple, ad library documentation, and existing academic efforts, we categorize mobile ad regulations into three types:
\begin{itemize}[noitemsep, topsep=1pt, partopsep=1pt, listparindent=\parindent, leftmargin=*]
\item \textit{Placement Regulation}: It regulates the size, location, and visibility of ads~\cite{google-banner,google-floating-banner,meta-policy,applovin-policy}. Common violations include oversized banners, ads that overlap content, or ads that are hidden behind or within interactive UI elements.
\item \textit{Interaction Regulation}: It specifies how ads should behave when users interact with them~\cite{google-interstitial,apple-guideline,betterads-standard}. For example, ads should not be displayed during app launch or exit, and full-screen ads must allow users to close them immediately~\cite{google-interstitial,apple-guideline,betterads-standard}.
\item \textit{Content Regulation}: It regulates the content of ads. Violations include harmful content and drive-by-download~\cite{google-content-policy,google-clickbait,google-malware,unity-policy}.
\end{itemize}
\autoref{tab:adpolicy} summarizes the three categories of ad regulations, along with references from platforms, ad networks, and prior research. We observe that both major platforms (Google and Apple) and popular ad networks (e.g. Unity, Meta Ads, AppLovin) enforce requirements across all three categories. In contrast, existing academic approaches only address subsets of these regulations, leaving notable gaps between industrial requirements and research efforts. To the best of our knowledge, \sys is the first academic approach that supports all three types of regulation.  
 


\begin{figure}[t!]
    \centering
    \includegraphics[width=0.9\columnwidth]{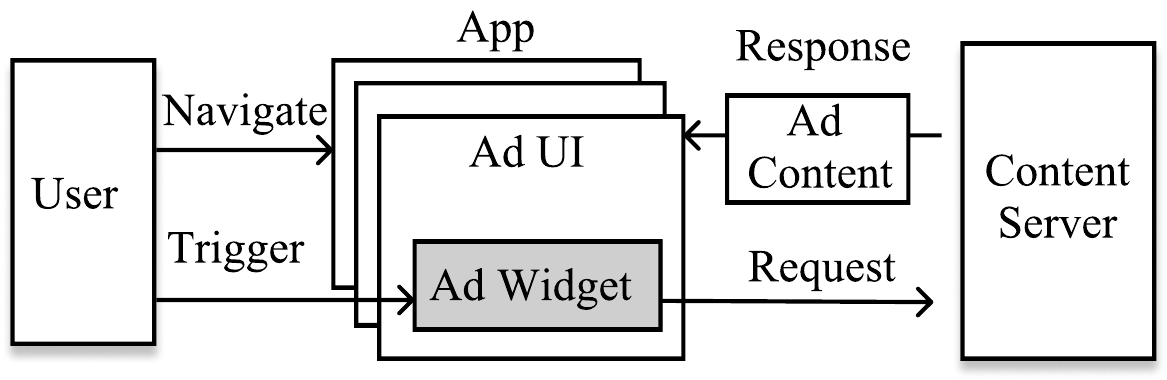}
    \caption{Mobile ad practice}
    \label{fig:Mobile Advertising Process}
    \vspace*{-1ex}
\end{figure}

To enforce mobile ad regulations effectively, it is essential to understand how ads are implemented and displayed in real-world mobile apps. \autoref{fig:Mobile Advertising Process} illustrates the typical pipeline for mobile ad practice: the user navigates through app UIs and interacts with an ad UI, which embeds ad widgets placed by the developer. When triggered, these widgets send ad requests to a content server and render the returned ad content.
In this pipeline, \textit{ad widgets play a central role as they serve as the entry point for ad rendering (Placement Regulation), mediate user interactions (Interaction Regulation), and deliver the final ad content (Content Regulation).}




\subsection{Ad Widgets Categorization}
 \label{subsec:empiricalstudy}

To better understand different types of ad widgets, we conduct a characteristic study on a dataset built from AndroZoo~\cite{allix2016androzoo}, a well-known and frequently updated collection of Android applications. We choose apps from AndroZoo based on three criteria: (1) the apps must be released or updated after 2022; (2) they need to include ad libraries, indicating they have mobile ads; (3) their download numbers from the app market must vary, as apps with different levels of popularity may use ads differently. 
After filtering AndroZoo using all these criteria, we randomly sample $100$ apps, which span across $35$ app categories, varying in popularity from $10,000$ to $100$ million downloads and last updated after 2022, well representing the latest advertising practices.

We then manually analyze each app to record the ad widgets.
Specifically, our manual analysis consists of both dynamic analysis and static analysis.
We first execute the apps on real devices, as ad libraries do not show ads on emulators to avoid fraud. We then explore the apps, navigate to all the ad UIs, and trigger the ad widgets.  During this process, we develop a label tool to record the screenshots of ad UIs and the ad content and dump the hierarchies of ad UIs.
Additionally, we decompile the apps using dex2jar~\cite{dex2jar}, and manually inspect the code of ad widgets.
Within our affordable efforts, we have found $250$ ad widgets in $100$ apps. ~\autoref{tab:adwidgets} summarized our observations.

\begin{itemize}[noitemsep, topsep=1pt, partopsep=1pt, listparindent=\parindent, leftmargin=*]
    \item \textit{Adview Widgets:} Accounting for $40.8\%$ ($102$) of our dataset, these widgets are shown using visible views created via ad libraries and are automatically triggered when apps are started. They support \textit{placement} and \textit{content} regulations. 
    \item  \textit{Popup Widgets:} Accounting for $24.4\%$ ($61$) of our dataset, these widgets are created using ad libraries and are usually triggered by user interaction, such as clicks. They support \textit{interaction} and \textit{content} regulations. 
    \item \textit{Native Widgets:}  Accounting for $34.8\%$ ($87\
    $) of the dataset, these widgets are developed in the app's native code, content provided by the app developer. They support all three categories of regulations.  
\end{itemize}




The diverse creation mechanism and behaviors of ad widgets pose challenges for existing static analysis that focuses mainly on ad library usage and motivates the design of \sys.

\begin{table}[t]
\caption{Characteristic study on ad widgets}
\label{tab:adwidgets}
\resizebox{\columnwidth}{!}{%
\begin{tabular}{@{}llll@{}}
\toprule
Ad widget & Source          & Occurences & Supported Regulation \\ \midrule
Adview    & Ad Library      & 102       & Placement, Content   \\
Popup     & Ad Library      & 61       & Interaction, Content \\
Native    & Native app code & 87       & Interaction, Placement, Content   \\ \bottomrule
\end{tabular}
}
\end{table}

\section{Overview of \sys}

\begin{figure*}[t!]
    \centering
    \includegraphics[width=\textwidth]{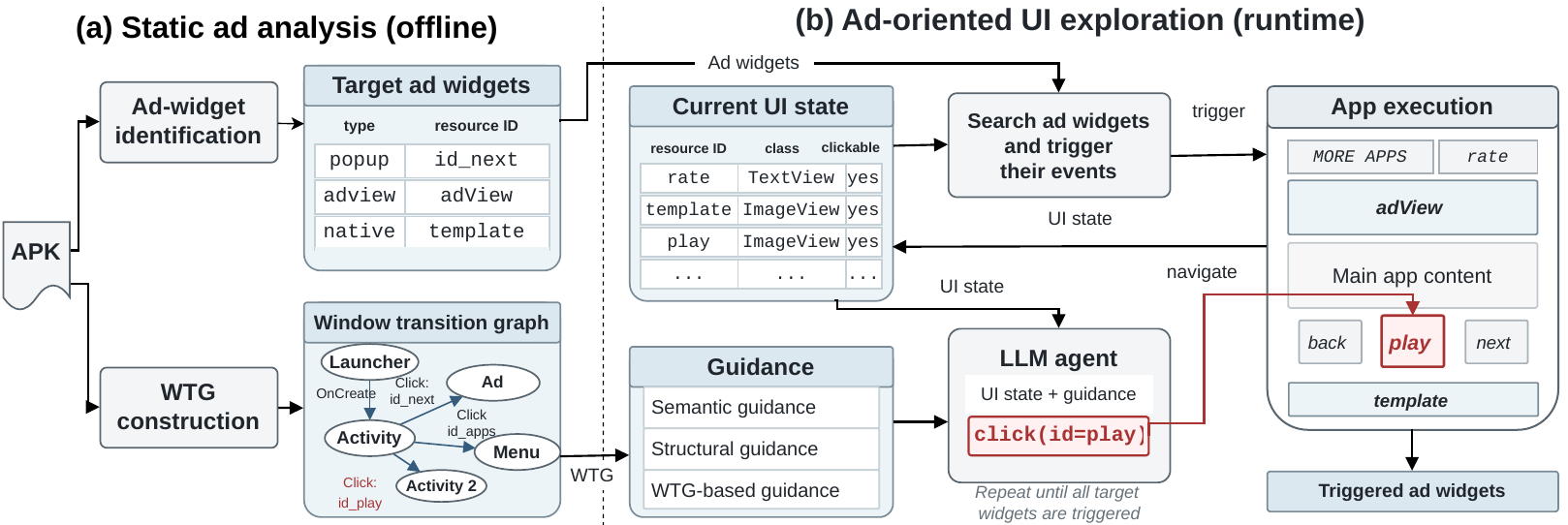}
    \caption{Overview of \sys}
        \vspace*{-1ex}
    \label{fig:overview}
\end{figure*}

\autoref{fig:overview} shows the overview of \sys, which consists of two main phases: static ad analysis (offline) and \uinav (runtime). 

In the offline phase, \sys takes the app's APK file as input to statically identify ad widgets based on their attributes/behaviors and constructs a WTG. 
Examples of the output ad widgets and WTG are provided in the figure.

In the runtime phase, \sys executes the app to navigate to ad UIs and trigger ad widgets.  
Specifically, for each UI currently being visited, \sys searches for ad widgets identified in static ad analysis based on resource-id (e.g., the widget with resource-id ``template'') and triggers them based on their events (e.g., ``click''). 
Once all ad widgets have been triggered in the current UI, \sys computes three types of exploration guidance (WTG-based guidance, semantic guidance, and structural guidance) that guide the LLM in choosing a UI widget to click, facilitating navigation to another UI for further exploration. 
For instance, as shown in \autoref{fig:overview}, \sys chooses the widget with resource-id ``play'' to navigate to the next UI (text highlighted in red).
This process repeats until all ad widgets of the app are triggered or resource limits are met. 
We next describe these two phases in detail.

\section{Static Ad Analysis}
Based on the observations from the characteristic study (Subsection~\ref{subsec:empiricalstudy}), in this section, we formally define ad widgets and the related concepts.  We then present the technique for ad widget identification and WTG construction. 

\subsection{Definitions of Ad Widgets}

\begin{definition}[UI Layout, Widget]
A UI layout $L(W)$ is a rooted tree structure, where each node $w(v, id, I) \in W$ represents a widget, with $v$ denoting the view class, $id$ being a unique layout/resource identifier, and $I = \{i_1, i_2, ..., i_n\}$ being a set of invocation events, where each invocation event $i$ represents a method call made on the widget object. $w$ is uniquely identified using both $L.id$ and $w.id$.
\end{definition}

\begin{figure}[t!p]
	\begin{lstlisting}
<LinearLayout>
    <com.google.android.ads.nativetemplates.TemplateView android:id="@+id/my_template">
    <ImageView android:id="@+id/template>
    <LinearLayout>
        <ImageView android:id="@+id/id_play> ...
    </LinearLayout>
    <com.google.android.gms.ads.adview android:id="@+id/adView"> ...
</LinearLayout>
\end{lstlisting}
    \vspace*{-1ex}
	\caption{Simplified layout}
	\label{code:layout}
	\vspace*{-1ex}
\end{figure}

\begin{figure}[t!p]
	\begin{lstlisting}
 import com.google.android.gms.ads.Adview;
 public void onCreate(Bundle bundle) {
    setContentView(R.layout.activity_list_rengtone);
    Adview adView = (Adview) findViewById(R.id.adView); 
    ...}\end{lstlisting}
        \vspace*{-1ex}
	\caption{Code snippet of an adview widget}
	\label{code:adview}
	\vspace*{-1ex}
\end{figure}



\begin{figure}[t!p]
	\begin{lstlisting}
import com.google.android.gms.ads.InterstitialAd;
import android.widget.Button;
public void onCreate(Bundle bundle) {
    InterstitialAd interstitialAd = new InterstitialAd(this);
    Button adButton = (Button) findViewById(R.id.b1);
    adButton.setOnClickListener(new View.OnClickListener(){
        public void onClick(View view) {
            interstitialAd.show();
        } }); ...}\end{lstlisting}
            \vspace*{-1ex}
	\caption{Code snippet of a popup widget}
	\label{code:popup}
	\vspace*{-1ex}
\end{figure}

\begin{figure}[t!p]
	\begin{lstlisting}
public boolean onOptionsItemSelected(MenuItem menuItem) {
    int itemId = menuItem.getItemId();
    if (itemId == R.id.apps) {
        startActivity(new Intent("android.intent.action.VIEW", Uri.parse("https://play.google.com/store/apps/developer?id=" + getString(R.string.developer_id_or_name))));} ...}\end{lstlisting}
            \vspace*{-1ex}
	\caption{Code snippet of a native widget}
	\label{code:custom}
	\vspace*{-1ex}
\end{figure}

\begin{definition}[Ad Widget]
An ad widget $w_{a}$ is a type of widget that developers use to place ads.
\end{definition}
For example, as shown in \autoref{code:layout}, the UI layout $L$ is \incode{R.layout.activity_list_ringtone}, which includes adview widgets (Line 2 and Line 8), popup widgets (Line 5), and native widgets (Line 3).



We next define three types of ad widgets based on their characteristics. 
\begin{definition}[Adview Widget]  
         An adview widget is a specialized view created by an ad library to show ads.
\end{definition}

 All ad libraries define a set of specialized view classes for showing ads. 
For example, the banner ad labeled ``Tranqull Luxury Spa Retreat'' at the bottom of the screenshot on the right side of \autoref{fig:overview} is created using the class \incode{com.google.android.gms.ads.Adview} in \autoref{code:adview}, with a view class defined in the AdMob library (Line 1), and the resource-id of the widget is specified as \incode{R.id.adView} (Line 4). 
Similarly, the ``Call number locator'' ad at the top of the screenshot is created using the class \incode{com.google.android.ads.nativetemplates.TemplateView} with the resource-id \incode{my_template}.

  


\begin{definition}[Popup Widget]
A popup widget has a method invocation that calls APIs from ad libraries to show ads.
\end{definition}

Unlike adview widgets, the view classes of popup widgets are chosen by the developers and 
do not necessarily come from the view classes of specific libraries.
Nonetheless, they share a common feature: interacting with the widget (usually via a click) will display an ad from the ad library.
For example, \autoref{code:popup} shows the code snippet that triggers a popup ad when a button is clicked. Specifically, the class of the popup widget $adButton$ is \incode{android.widget.Button} and its resource-id is defined as  \incode{R.id.b1} (Line 5). This widget is bound to a click event that calls the \incode{show()} method to display the ad created using \incode{com.google.android.gms.ads.InterstitialAd} (Line 6-8).

\begin{definition}[Native Widget]
A native widget has a method invocation that redirects users to the advertised content in either the app market or a website.
\end{definition}

\autoref{code:custom} shows a native widget that will trigger an ad when users click the \incode{menuItem} of an \incode{OptionalMenu}. 
The code checks if the \incode{menuItem} with the resource-id  \incode{R.id.apps} is selected (Line 3), and if so, it starts an Activity with an Intent pointing to the developer's Google Play store page (Line 4).
Similar to popup widgets, the view classes of native widgets are chosen by the developers and thus do not have specific patterns. Nevertheless, they share a common feature: once triggered, they redirect users to download their promoted apps or visit their promoted product's website.

\begin{figure*}[t]
\centering
\small
\setlength{\abovecaptionskip}{2pt}

\begin{minipage}[b]{0.28\textwidth}
    \centering
\AxiomC{$w(v,id,I) \in L(W) $}
\UnaryInfC{$v \in View(AdLib)$}
\alwaysSingleLine
\LeftLabel{$\mathcal{R}_a: $}
\UnaryInfC{$w$ is an adview widget}
    \DisplayProof
\end{minipage}%
\hfill
\begin{minipage}[b]{0.28\textwidth}
    \centering
\AxiomC{$w \in L(W) $} 
\UnaryInfC{$\exists i \in I: \text{target}(i) \in \text{API}$}
\alwaysSingleLine
\LeftLabel{$\mathcal{R}_p: $}
\UnaryInfC{$w$ is a popup widget}
    \DisplayProof
\end{minipage}%
\hfill
\begin{minipage}[b]{0.40\textwidth}
    \centering
\AxiomC{$w \in L(W) $}
\UnaryInfC{$\exists i \in I: \text{hasIntent}(i, \text{URL} \cup \text{Market})$}  
\alwaysSingleLine
\LeftLabel{$\mathcal{R}_n: $}
\UnaryInfC{$w$ is a native widget}
    \DisplayProof
\end{minipage}

\caption{Inference rules for UI exploration.}
\label{proof:adwidgets}
\end{figure*}

\subsection{Ad Widget Identification}
\label{subsec:adWidgetIdentification}



Based on the ad widget definitions, we develop a novel static analysis framework that identifies ad widgets in an Android app. 
Specifically, \sys models a widget as $w(v, id, I)$, where the view class $v$ and the method invocations $I$ are defined in an app's code, and the resource-id $id$ is separately specified in the app's UI layout files.
Given that a widget’s view class and resource-id are defined via different Android framework APIs, and its method invocations correspond to the set of method calls on the program variable that represents the widget, \sys employs a novel dataflow analysis combining application code and UI layout information to identify each widget’s view class $v$ and associate $v$ with its resource-id $id$ and method invocations $I$.
Existing approaches~\cite{yang2015static,xiao2019iconintent,gator3} are restricted to detecting behaviors tied to built-in event handlers, overlooking many widget behaviors such as \incode{MenuItem} clicks and \incode{Adaptor} events. 
To overcome this limitation, \sys extends the analysis to \textit{all method invocations on widgets}, enabling broader coverage of ad-related behaviors. 
Since this inevitably introduces unrelated behaviors that do not trigger ads, \sys applies inference rules derived from ad widget definitions to evaluate $v$ and $I$, filtering out irrelevant cases and retaining only the qualified ad widgets.
The resource-ids $id$ are later used to guide ad-oriented UI exploration to look for the UIs that contain the ad widgets. 
We next describe the details of this framework, which consists of two phases:


\noindent\textbf{Dataflow Analysis.}
In the first phase, \sys associates each widget $w(v, id, I)$ with its view class $v$, resource-id $id$,  and method invocations $I$. 
Specifically, \sys first scans the application code for method calls \incode{findViewById(w.id)}, which returns a view object bound to a variable $x$, establishing the mapping $x \rightarrow w.id$.
Then,  \sys conducts a type analysis on $x$ to resolve the type of the object referenced by $x$ (e.g., \incode{(AdView) x}), thus creating the mapping $x \rightarrow v$.
Finally, \sys performs a reference analysis to find all aliases of $x$ and collects every method call made on them, thus obtaining the mapping $x \rightarrow I$. 
This process computes the complete attributes for each widget $w(v, id, I)$.

\noindent\textbf{Inference Rule Checking.}
In this phase, \sys applies a set of inference rules to identify an ad widget based on its view class $v$ and invocations $I$. 
These rules, shown in \autoref{proof:adwidgets}, are derived based on the definition of each type of ad widget introduced earlier.
\begin{itemize}[noitemsep, topsep=1pt, partopsep=1pt, listparindent=\parindent, leftmargin=*]
    \item \textit{$\mathcal{R}_a$:} $w$ is an \emph{adview widget} if its view class $v$ belongs to the set of ad library view classes.
    \item \textit{$\mathcal{R}_p$:} $w$ is a \emph{popup widget} if any invocation $i \in w.I$ calls an ad library API, i.e., $\exists i \in w.I: \text{target}(i) \in \text{API}(\text{AdLib})$.
    \item \textit{$\mathcal{R}_n$:} $w$ is a \emph{native widget} if any invocation $i \in w.I$ launches an \incode{Intent} to an app market or website, i.e., $\exists i \in w.I: \text{hasIntent}(i, \text{URL} \cup \text{Market})$.
\end{itemize}
For $\mathcal{R}_a$ and $\mathcal{R}_p$, we curate the ad libraries list released by existing studies~\cite{xinyu2023andetect,ma2016libradar}. We further expand this list by applying ad library detection tools~\cite{ma2016libradar} to the dataset in Section~\ref{sec:motivationionalstudy}, ultimately obtaining a total of $41$ ad libraries. 
Note that all inference rules are necessary conditions but not sufficient conditions for identifying ad widgets. For example, a widget can trigger a click event to open a privacy policy webpage (i.e., having Intent to website) but may not be a native widget. Precisely detecting ad widgets through static analysis is challenging and often impractical due to the diverse coding practices employed by developers. 
Implementing these rules, however, enables \sys to capture all potential ad widgets from the dataflow facts. 
Any irrelevant widgets can then be easily filtered out during the subsequent dynamic UI exploration phase.

\subsection{WTG Construction}
\label{sec:wtgAnalysis}

Ad widget identification identifies the set of candidate ad widgets.
But to effectively trigger them at runtime, we must also reason about how to reach the ad UIs in which these widgets are embedded (such as navigating from the main activity to an ad UI that shows an ad when a button is clicked). 
\sys models such navigation across UIs with a WTG and present its formal definition as follows.

\noindent\textbf{Definition of WTG.}
A WTG is defined as a directed graph: $G = (Win, E, \epsilon)$ where:
\begin{itemize}[noitemsep, topsep=1pt, partopsep=1pt, listparindent=\parindent, leftmargin=*]
\item $Win$ denotes a set of nodes, each representing a UI (also called a window) in the app. A window can be an \emph{Activity}, \emph{Dialog}, \emph{Menu}, or \emph{Fragment}.
\item $E \subseteq Win \times Win$ denotes a set of directed edges, each representing a transition from a source UI to a target UI.
\item $\epsilon : E \rightarrow w(v, id, I)$ denotes an edge-labeling function mapping each transition to the widget that triggers it.
\end{itemize}
\noindent\textbf{Static Construction Process.}
\sys constructs $G$ in three steps.
 First, \sys identifies candidate windows $Win$ by analyzing specific API invocations such as \incode{setContentView}, \incode{Dialog.show}, and \incode{MenuInflater.inflate}. For fragments, \sys tracks \incode{FragmentTransaction.add/replace} calls to recover dynamically attached UIs.   Next, for each method containing navigation-related calls (e.g., \incode{startActivity} and \incode{FragmentTransaction}), \sys adds edges $(Win_s, Win_t) \in E$ between the source window $Win_s$ and the resolved target $Win_t$. 
 Finally, \sys computes $\epsilon(e)$ by linking each transition edge $e \in E$ back to the widget $w$ that invokes the transition in its method invocation $w.I$.

\begin{algorithm}[t]
\small
\caption{Ad-Oriented UI Exploration}
\label{alg:LLM}
\KwIn{$maxStep$\;\\
Static Ad Analysis: $wtg$, $adWidgets$\;\\
Placement Information: $metadata$, $knowledgeBase$\;\\
Runtime UI State: $UI$\;
}
$steps \gets 0$\;
$actionHistory \gets \emptyset$\;
$semanticGuid \gets \text{SummarizeFunctionality}(metadata)$\;

\While{$adWidgets \neq \emptyset$ \textbf{and} $steps < maxStep$}{
    $adWidgets \gets \text{Trigger}(adWidgets, UI)$\;
    \If{$adWidgets = \emptyset$}{\textbf{break}\;}
    
    $wtgGuid \gets \text{GetShortestPath}(UI, adWidgets, wtg)$\;
    $structuralGuid \gets \text{FindSimilarUIs}(knowledgeBase, UI)$\;
    $guidance \gets \{wtgGuid, semanticGuid, structuralGuid\}$\;
    $verifiedUI \gets \text{VerifyUI}(UI)$\;
    $w \gets \text{SelectWidget}(guidance, verifiedUI, actionHistory)$\;
    $newUI \gets \text{Navigate}(UI, w)$\;
    $outcome \gets \text{ReflectAction}(UI, newUI, w)$\;
    $actionHistory \gets \text{UpdateHistory}(actionHistory, w, outcome)$\;
    $UI \gets newUI$\;
    $steps \gets steps + 1$\;
}
\end{algorithm}
\section{Ad-Oriented UI Exploration}
We next describe how \sys instructs LLMs to digest static analysis results and ad widget placement strategies to guide UI exploration towards ad UIs and trigger ad widgets.

\noindent\textbf{Workflow.}
Algorithm~\ref{alg:LLM} presents the workflow of \sys's \uinav technique.
Given the static ad analysis results, ad placement information, and current UI state, \sys initializes a short-term action history and generates semantic guidance from the app metadata (Lines 1--3).
At each step, \sys triggers visible ad widgets and updates the remaining targets (Lines 4--8).
It then computes WTG-based and structural guidance and combines them with the semantic guidance (Lines 9--11).
To ground action selection in runtime evidence, \sys filters the current UI to retain only visible and interactable widgets (Line 12).
The LLM selects a widget using the combined guidance, verified UI, and action history, and then executes the action (Lines 13--14).
Afterward, \sys compares the pre-action and post-action UI states and records the outcome, allowing subsequent steps to avoid ineffective actions and revise the exploration strategy (Lines 15--16).
Finally, it updates the UI state and step count (Lines 17--18).
This process continues until all identified ad widgets are triggered or the exploration budget is exhausted (Line 4).

\begin{figure}[t]
    \centering
   
\begin{promptbox}[Ad-Oriented UI Exploration Prompt]
\footnotesize
Act as an Android UI testing expert.

\smallskip
\noindent\fbox{\textbf{Semantic Guidance}} \\[2pt]
\textit{Listen to Taj's songs and explore high-quality audio content on the app Taj Music \& Audio. Navigate through the app and click any buttons or links that lead to other apps or advertising content}

\smallskip
\noindent\fbox{\textbf{WTG-Based Guidance}} \\[2pt]
Based on the window transition graph, the current activity \textit{Activity1} can transfer to the following activities: \\
\quad \textit{Activity2} \textit{visited} with cost \textit{1} via widget \textit{resource-id=id\_about} to \textit{click} \\
\quad \textit{OptionsMenu} \textit{not-visited} with cost \textit{1} via widget \textit{resource-id=id\_apps} to \textit{click} \\
\quad \textit{$\langle$Target Activity$\rangle$} \textit{$\langle$if\_visited$\rangle$} with cost \textit{$\langle$cost$\rangle$} via widget (\textit{$\langle$Attributes$\rangle$}) to \textit{$\langle$Event$\rangle$}

\smallskip
\noindent\fbox{\textbf{Structural Guidance}} \\[2pt]
Below are $\langle k \rangle$ most similar UIs of the current UI. \\
In UI-1 we have 2 UI widgets, namely: \\
\quad \textit{index -0: (resource-id: llFeatureApps, text: Featured Apps)} to \textit{click} \\
\quad \textit{index -1: (resource-id: llMoreApps, text: More Apps)} to \textit{click} \\
\quad In this UI, \textit{index -1: (resource\_id: llMoreApps, text: More Apps)} can lead to advertising content... \\
In UI-k we have $\langle$\textit{Number of Widgets}$\rangle$ UI widgets, namely: \\
\quad \textit{index -0: $\langle$Widget\_Attributes$\rangle$} to \textit{$\langle$Event$\rangle$} \\
\quad \textit{index -1: $\langle$Widget\_Attributes$\rangle$} to \textit{$\langle$Event$\rangle$}... \\
\quad In this UI, \textit{index -n: $\langle$Widget\_Attributes$\rangle$} can lead to advertising content \\[2pt]
Currently, we have 12 UI widgets, namely: \\
\quad \textit{index -0: view1 (resource-id=apps, text=MORE APPS)} to \textit{click} \\
\quad \textit{index -1: view2 (resource-id=adView)} to \textit{click}... \\[4pt]
Please choose only one UI element with its index such that the element can bring us closer to our test target. If none of the UI elements can do so, respond with index-none.
\end{promptbox}

    \caption{Example prompt for three types of domain guidance.}
    \label{fig:promptguidance}
    \vspace*{-1ex}
\end{figure}


\subsection{Domain Guidance for UI Exploration}
\label{subsec:domainguidance}
\sys computes three types of domain guidance for the LLM. The guidance prompt is summarized in \autoref{fig:promptguidance}.

\noindent\textbf{WTG-Based Guidance.}
In principle, the static analysis results allow us to compute optimized transition paths in the WTG to reach ad UIs. However, as noted in prior work~\cite{liu2022promal,Gator}, statically constructed WTGs often contain infeasible UI transitions. In contrast, LLMs have shown strong capabilities in understanding UIs and code~\cite{chatgpt2,nam2024using,ran2024guardian}. Therefore, \sys provides WTG information to the LLM to improve its reasoning and planning during UI exploration.

Since a WTG is represented as a graph while LLM input is a sequence of text tokens, \sys converts the WTG into textual guidance by applying Dijkstra's algorithm~\cite{dijkstra1959note} to compute the shortest-path distance from the current UI to other reachable ad UIs. This shortest-path information helps the LLM optimize navigation, enabling more efficient and cost-effective exploration.

\noindent\textbf{Semantic Guidance.}
\label{sec:explorationTask}
Developers earn revenue when ad-library ads (i.e., adview and popup ads) are viewed by users (cost per impression~\cite{cpm}) or clicked by users (cost per click~\cite{cpc}). To maximize revenue, developers tend to place ads in UIs that target users are most likely to visit, which are often close to app's main functionality. Moreover, prior work~\cite{wen2024autodroid,ran2024guardian} shows that providing explicit exploration guidance substantially improves the ability of LLMs to reach target UIs. Based on these observations, \sys generates semantic guidance that encourages the LLM to explore the app in a way that resembles target-user behavior.

Specifically, we first crawl app metadata from the app market, including the app category, name, and description. \sys then prompts the LLM to summarize the app's main functionality based on this metadata. For example, ``\textit{Listen to Taj's songs and explore high-quality audio content.}'' is the semantic guidance generated for the ``Taj Music \& Audio'' app in \autoref{fig:overview}. In addition, our empirical study shows that native widgets often appear in menus and pop-up dialogs near app entry or exit. To encourage the LLM to consider such widgets, \sys appends the instruction: ``Navigate through the app and click any buttons or links that lead to other apps or advertising content.''



\noindent\textbf{Structural Guidance.}
Certain types of ad widgets are placed in UIs that are not directly related to the app's major functionality but are still likely to entice user engagement, such as the ``MORE APPS'' menu item in \autoref{fig:overview}'s app screen. 
Therefore, it is difficult to describe these ad widget placements effectively as part of the semantic exploration guidance.
To address this, we find that the strategies for placing ad widgets are often similar among apps with similar UI designs. 
For example, native widgets are typically placed next to or inside the menu if the app has a menu on top of a UI. 
To capture these similarities, \sys employs a RAG-based prompting approach, which enhances the LLM's capabilities in finding structural ad widget placements using similar ad UIs.
Specifically, \sys builds a knowledge base of representative ad UIs and their structural patterns. During exploration, it retrieves UIs similar to the current screen and uses their ad-widget placements as structural guidance for the LLM.


\subsection{Grounded UI Reasoning} 

While LLMs are effective at understanding UI content, they often hallucinate during planning and automation in complex mobile UIs~\cite{wen2024autodroid,ran2024guardian}. 
To address this challenge, \sys incorporates two techniques that ground the LLM's reasoning in observable runtime evidence, drawing on established principles from the ReAct~\cite{yao2022react} and self-refinement~\cite{madaan2023self} paradigms.

\noindent\textbf{Action Reflection.} After each exploration step, \sys prompts the LLM to compare the pre-action and post-action UI states and explicitly judge whether the intended navigation occurred. If the LLM determines that the UI remained unchanged or transitioned to an unexpected screen, it marks the previous action as \emph{failed} and appends this outcome to a short-term action history maintained across steps. This reflective feedback serves two purposes: (1) it prevents the agent from re-selecting the same ineffective action in subsequent steps, breaking action repetition loops that are a well-documented failure mode in LLM-based mobile agents~\cite{wen2024autodroid}, and (2) it enables the LLM to adaptively revise its exploration strategy. For example, by choosing an alternative navigation path suggested by the WTG guidance when a direct path proves infeasible at runtime. 

\noindent\textbf{UI State Verification.} Before presenting the current UI to the LLM for action selection, \sys performs a verification pass that filters the raw UI hierarchy to retain only genuinely interactable elements. Specifically, we remove widgets that are marked non-clickable, obscured by overlapping views, or located outside the visible screen bounds. This verified, distilled UI representation reduces the LLM's action space and mitigates a key source of hallucination: the tendency to select plausible-looking but non-functional UI elements from a cluttered view hierarchy. By narrowing the candidate set to verified, actionable widgets, we improve both the accuracy of the LLM's action selection and the efficiency of the overall exploration, as fewer steps are wasted on invalid interactions.

Together, reflection and verification form a lightweight feedback loop that allows the LLM to revise its decisions based on runtime observations, rather than relying solely on one-step reasoning from the current screenshot or UI hierarchy. The action-reflection prompt is summarized in \autoref{fig:promptactionreflection}.

\begin{figure}[h]
    \centering

\begin{promptbox}[Action Reflection Prompt]
\footnotesize
\textbf{System:} You are a mobile UI testing agent. Your task is to determine if the previous action successfully navigated to a new UI state.

\smallskip
\textbf{Input:}
\begin{itemize}[leftmargin=*, nosep]
    \item \texttt{previous\_UI}: The UI state before the action was executed.
    \item \texttt{current\_UI}: The UI state after the action was executed.
    \item \texttt{action}: The action that was performed (e.g., \texttt{click(widget\_id)}, \texttt{scroll(direction)}).
    \item \texttt{action\_history}: A list of recent actions and their outcomes.
\end{itemize}

\smallskip
\textbf{Prompt:} \\
You have just performed the action \texttt{\{action\}} on the mobile app UI.

Below is the UI state \textbf{before} the action: 
\texttt{\{previous\_UI\}}

Below is the UI state \textbf{after} the action: 
\texttt{\{current\_UI\}}

Here is the history of recent actions and their outcomes: 
\texttt{\{action\_history\}}

Based on the above, answer the following:
\begin{enumerate}[nosep]
    \item Did the UI state change after the action? (Yes/No)
    \item If yes, describe briefly what changed (e.g., new screen, dialog appeared, list scrolled).
    \item If no, hypothesize why the action may have failed (e.g., widget was non-interactable, action was invalid for the current state).
    \item Should this action be \textbf{avoided} in subsequent steps? (Yes/No)
    \item Suggest an alternative action if the previous one failed.
\end{enumerate}

\end{promptbox}
    \vspace*{-1ex}
    \caption{Prompt template for Action Reflection.}
    \label{fig:promptactionreflection}
    \vspace*{-1ex}
\end{figure}

\subsection{Rule-based Detection of Ad Regulation Violation}
\label{subsec:violationdetection}
After triggering ad widgets and retrieving their content, \sys applies a set of detection rules to identify violations of ad regulations.  
In this work, we focus on the most impactful types of violations across three ad regulation categories that negatively affect user experiences or pose risks to user privacy.

 We consider two types of  \textbf{Placement Violations}:
\begin{itemize}[noitemsep, topsep=1pt, partopsep=1pt, listparindent=\parindent, leftmargin=*]
\item \textit{Oversized Ads.} Ads that occupy more than 30\% of the screen are considered oversized~\cite{betteradstandards}.
\sys detects such ads by checking the \incode{Bounds} attribute of adview and native widgets.
\item \textit{Overlapped Ads.} Ads that obstruct or overlap with other interactive app content are considered violations~\cite{google-banner}. \sys detects such ads by finding overlaps between the bounds of ad widgets and other clickable or input widgets.
\end{itemize}
 We consider two types of \textbf{Interaction Violations}:
\begin{itemize}[noitemsep, topsep=1pt, partopsep=1pt, listparindent=\parindent, leftmargin=*]
\item \textit{Disruptive Ads.} These include popup ads shown immediately when the app loads or before the app exits. 
\sys detects such ads by checking whether there is an interstitial ads after loading or before exiting the app.   
\item \textit{Unskippable Ads.} Full-screen ads that cannot be closed promptly~\cite{betterads-standard,google-interstitial}.  
After triggering a popup widget, \sys checks whether it gets stuck in an ad library's activity for two consecutive actions. 
The rationale is that if the ad is skippable, the LLM should be able to close it easily.
\end{itemize}
 We also consider two types of \textbf{Content Violations}:
 \begin{itemize}[noitemsep, topsep=1pt, partopsep=1pt, listparindent=\parindent, leftmargin=*]
\item \textit{Malvertising.} Ads can be exploited by attackers to distribute malware~\cite{ma2024careful} or direct users to malicious websites~\cite{rastogi2016these}.  Some ads redirect users to download pages without clear consent.  
\sys records the landing URL after the ad is triggered to detect such behavior. These apps and URLs are then scanned using VirusTotal to assess their maliciousness. 
We consider an app or URL malicious if it is flagged by at least three VirusTotal engines, following common practice.
\item \textit{Clickbait Ads.} Ads that lure users to click using exaggerated claims or deceptive images. 
To detect these ads, \sys prompts a multimodal LLM  to ask whether the ad content is related to clickbait. 
\end{itemize}

Note that although the above-mentioned violations do not cover all real-world cases, \sys can be easily tailored based on ad widgets and their attributes to support different policies and specifications.


\section{Evaluation}
We evaluate the effectiveness of \sys in detecting mobile ads using real-world apps.
Specifically, we aim to answer the following research questions:
\begin{itemize}[noitemsep, topsep=1pt, partopsep=1pt, listparindent=\parindent, leftmargin=*]

    \item \textbf{RQ1:} How effectively and efficiently can \sys detect ad widgets, and thereby detect ad regulation violations? 
    \item \textbf{RQ2:} How effectively can \sys's static ad analysis help detect ad widgets? 
        \item \textbf{RQ3:} How effectively can \sys's ad-oriented UI exploration help detect ad widgets?    
        \item \textbf{RQ4:} How effectively can \sys detect ad regulation violations in the wild?
\end{itemize}

\subsection{Evaluation Setup}

\noindent\textbf{Dataset.}
Following the strategy described in Subsection~\ref{subsec:empiricalstudy}, we sample another dataset of $100$ apps, representative of the latest advertising practices, to evaluate \sys. 
This dataset contains $271$ ad widgets, including $122$ AdView, $51$ Popup, and $98$ Native widgets.

\noindent\textbf{Metric.} 
We measure the effectiveness of ad widget detection using \textit{detection rate}. 
Specifically, an ad widget is considered detected if it, or any of its children UI widgets, is triggered during exploration. 
We define this formally as follows.
Suppose the set of triggered widgets during exploration for an approach is $W_t$. An ad widget $w_a$ is marked as detected if it meets one of the following conditions for $w_t \in W_t$:
\begin{itemize}[noitemsep, topsep=1pt, partopsep=1pt, listparindent=\parindent, leftmargin=*]
    \item One of $w_a$'s children UI widgets has a resource ID that matches the resource ID of $w_t$.
    \item $w_t$ is nested within $w_a$, where $w_t$'s bounding box is entirely enclosed within the bounding box of $w_a$.
\end{itemize} 
We then obtain the detection rate by computing the ratio of detected ad widgets to the total number of ad widgets.
Our goal is to detect more ad widgets in less time.  
To further evaluate efficiency, we measure the average time required to detect an ad widget, which we denote as \textit{Per-Widget Detection Latency (PWDL)}.  
Formally, this is defined as
\[
\text{Per-Widget Detection Latency} = \frac{T_{\text{last}} - T_{\text{start}}}{N_{\text{detected}}},
\]
where $T_{\text{start}}$ is the start timestamp of the exploration, $T_{\text{last}}$ is the timestamp when the last ad widget is detected, and $N_{\text{detected}}$ is the total number of ad widgets detected.


\noindent\textbf{Implementation.} 
\sys's static ad analysis is implemented in Java based on Soot~\cite{vallee2010soot} and FlowDroid~\cite{arzt2014flowdroid}.  
\sys's \uinav is implemented in Python. 
Since ad libraries implement emulator detection mechanisms to prevent fraudulent ad traffic, we conduct the UI exploration on real devices. 
Specifically, we use two smartphones with Android 11 for our evaluation.
We use GPT-5 as the underlying LLM model. 

\noindent\textbf{Baselines.} 
We compare \sys with baselines from these two aspects: ad detection~\cite{ma2024careful,liu2020maddroid,cai2023darpa} and guided UI exploration~\cite{ran2024guardian,lai2019goal}. 
\begin{itemize}[noitemsep, topsep=1pt, partopsep=1pt, listparindent=\parindent, leftmargin=*]
\item \textit{Text-Based Ad Detection:} AdGPE~\cite{ma2024careful} identifies ad content by detecting common call-to-action keywords and slogans, such as ``Install'', ``Play'', and ``Go''.
\item \textit{UI-Based Ad Detection:} MadDroid~\cite{liu2020maddroid}, FraudDroid~\cite{dong2018frauddroid}, and AdRambler~\cite{zhao2023mobile} adopt similar techniques to detect ads by targeting specific view classes (e.g., \texttt{AdView}, \texttt{WebView}, \texttt{ImageView}). Although none of these tools are open-sourced, we implement MadDroid based on the paper.
\item \textit{Vision-Based Ad Detection:} DARPA~\cite{cai2023darpa} uses trained object detection model (e.g., YoloV5) to identify visual UI patterns indicating ad content.
\item \textit{Static Analysis-Guided UI Exploration:} GoalExplorer~\cite{lai2019goal} statically locates a target (e.g., an API call, an activity) in the WTG and computes the shortest path to it, which is then used to guide a breadth-first search-based UI exploration.
\item \textit{LLM-Guided UI Exploration:} Guardian~\cite{ran2024guardian} and Autodroid\cite{wen2024autodroid} both instruct LLM with exploration guidance and let LLM plan UI exploration. We only compare with Guardian as Autodroid is not fully open-sourced.
\end{itemize}
We also set a control group using Android's Monkey~\cite{monkey}, implementing a depth-first search strategy to randomly explore different UIs. 
DARPA only provides an object detection model to detect ads without being integrated into UI exploration, so we manually record screenshots of all ad widgets in our dataset, apply the object detection model to them, and feed the results back to DARPA. 

\begin{table*}[t]
\caption{Comparison with baselines on ad widget detection and regulation violation detection}
\label{tab:overall}
\centering
\begin{tabular}{@{}lrrrrrlllr@{}}
\toprule
\multicolumn{1}{c}{\multirow{2}{*}{\textbf{Approaches}}} & \multicolumn{4}{c}{\textbf{Ad Widget Detection(\%)}}                                                                                             & \multicolumn{1}{c}{\multirow{2}{*}{\textbf{PWDL (s)}}} & \multicolumn{4}{c}{\textbf{Ad Regulation Violation Detection(\#)}}                                                                                            \\ \cmidrule(lr){2-5} \cmidrule(l){7-10} 
\multicolumn{1}{c}{}                                     & \multicolumn{1}{c}{\textbf{Adview}} & \multicolumn{1}{c}{\textbf{Popup}} & \multicolumn{1}{c}{\textbf{Native}} & \multicolumn{1}{c}{\textbf{Micro Avg.}} & \multicolumn{1}{c}{}                                   & \multicolumn{1}{c}{\textbf{Placement}} & \multicolumn{1}{c}{\textbf{Interaction}} & \multicolumn{1}{c}{\textbf{Content}} & \multicolumn{1}{c}{\textbf{Total}} \\ \midrule
Monkey~\cite{monkey}                                     & 18.85                               & 45.10                              & 19.39                               & 23.99                            & 13.18                                                  & 26                                     & 21                                       & 28                                   & 75                                 \\
MadDroid~\cite{liu2020maddroid}                          & 26.23                               & 47.06                              & 15.31                               & 26.20                            & 15.20                                                  & 28                                     & 14                                       & 32                                   & 74                                 \\
DARPA~\cite{cai2023darpa}                                & 22.13                               & 0.00                               & 0.00                                & 9.96                             & -                                                      & 0                                      & 0                                        & 23                                   & 23                                 \\
AdGPE~\cite{ma2024careful}                               & 28.69                               & 23.53                              & 16.33                               & 23.25                            & 12.52                                                  & 28                                     & 12                                       & 34                                   & 74                                 \\
GoalExplorer~\cite{lai2019goal}                          & 40.16                               & 39.22                              & 25.51                               & 34.69                            & 10.95                                                  & 31                                     & 20                                       & 48                                   & 99                                 \\
Guardian~\cite{ran2024guardian}                          & 13.93                               & 37.25                              & 16.33                               & 19.19                            & 21.38                                                  & 19                                     & 19                                       & 24                                   & 62                                 \\
\rowcolor{gray!15} 
\textbf{\sys}                                            & \textbf{68.85}                      & \textbf{58.82}                     & \textbf{51.02}                      & \textbf{60.52}                   & \textbf{10.00}                                         & \textbf{43}                            & \textbf{27}                              & \textbf{63}                          & \textbf{133}                       \\ \bottomrule
\end{tabular}
\end{table*}

\subsection{RQ1: Overall Effectiveness}


\noindent\textbf{Ad Widget Detection.} 
As shown in \autoref{tab:overall}, \sys effectively detects $68.85\%$ of adview widgets, $58.82\%$ of popup widgets, and $51.02\%$ of native widgets.
\sys achieves a micro-average detection rate of $60.52\%$, surpassing other ad detection methods by $36.53$ percentage points over random-based Monkey, $37.27$ percentage points over text-based AdGPE, $34.32$ percentage points over UI-based MadDroid, and $50.55$ percentage points over vision-based DARPA.
Moreover, \sys's micro-average detection rate outperforms the static analysis-guided UI exploration approach GoalExplorer by $25.83$ percentage points and the LLM-guided UI exploration approach Guardian by $41.33$ percentage points.
This demonstrates the effectiveness of the \sys's novel approach, which harnesses LLMs to unify static ad analysis and dynamic UI exploration with ad widget placement strategies and knowledge of ad UIs.

\noindent\textbf{Ad Regulation Violation Detection.} 
Once ad widgets are detected and triggered, violation detection rules can be applied based on their attributes (e.g., UI hierarchies, screenshots). To ensure fairness, we apply the same set of rules described in Subsection~\ref{subsec:violationdetection} to the ad widgets detected by each approach. 
As shown in \autoref{tab:overall}, \sys identifies a total of $133$ ad regulation violations, the highest among all approaches.  In comparison, Monkey detects $75$, MadDroid $74$, AdGPE $74$, Guardian $62$, and DARPA only $23$. The most effective existing approach, GoalExplorer, detects $99$ violations, while \sys achieves $34.34\%$ more. 
These results demonstrate the superior performance of AdWise in detecting ad
regulation violations, which directly come from \sys's higher effectiveness in ad widget detection.


\noindent\textbf{Efficiency.}  
\autoref{tab:overall} also shows the average time to trigger each ad widget for each approach.
We can find that \sys only spent $10.00$ seconds in triggering each ad widget, making it $24.12\%$ faster than Monkey, $34.21\%$ faster than MadDroid, $20.13\%$ faster than AdGPE, $8.68\%$ faster than GoalExplorer, and $53.23\%$ faster than Guardian.
These results demonstrate the superior efficiency of \sys in detecting ad widgets.



\subsection{RQ2: Static Ad Analysis}
\sys's static ad analysis can be incorporated with any other UI exploration approaches to detect ads. 
In this RQ, we evaluate the effectiveness of \sys's static ad analysis by measuring how its results enhance \sys and baselines for detecting ad widgets, as well as how effectively it can detect ad widgets alone.
\begin{table*}[t] 
\caption{The performance gains achieved by integrating \sys's static ad analysis. The results compare the original approach against the version enhanced with Static Analysis (+SA)}
\label{tab:sa}
\centering
\small
\setlength{\tabcolsep}{4pt} 
\begin{tabular}{@{}l c c c c c c c c@{}}
\toprule
\multirow{2}{*}{\textbf{Approaches}} & \multicolumn{2}{c}{\textbf{Adview}} & \multicolumn{2}{c}{\textbf{Popup}} & \multicolumn{2}{c}{\textbf{Native}} & \multicolumn{2}{c}{\textbf{Micro Avg.}} \\
\cmidrule(lr){2-3} \cmidrule(lr){4-5} \cmidrule(lr){6-7} \cmidrule(lr){8-9}
& Org. & +SA & Org. & +SA & Org. & +SA & Org. & +SA \\
\midrule

Monkey~\cite{monkey} 
& 18.85 & \textbf{36.07} & 45.10 & \textbf{47.06} & 19.39 & \textbf{21.43} & 23.99 & \textbf{32.84} \\

MadDroid~\cite{liu2020maddroid} 
& 26.23 & \textbf{39.34} & 47.06 & \textbf{47.06} & 15.31 & \textbf{18.37} & 26.20 & \textbf{33.21} \\

AdGPE~\cite{ma2024careful} 
& 28.69 & \textbf{36.07} & 23.53 & \textbf{39.22} & 16.33 & \textbf{18.37} & 23.25 & \textbf{30.26} \\

Guardian~\cite{ran2024guardian} 
& 13.93 & \textbf{36.89} & 37.25 & \textbf{52.94} & 16.33 & \textbf{22.45} & 19.19 & \textbf{34.69} \\

\sys 
& 36.89 & \textbf{68.85} & 58.82 & \textbf{58.82} & 33.67 & \textbf{51.02} & 39.85 & \textbf{60.52} \\

\midrule
Static Analysis (Only) 
& \multicolumn{2}{c}{81.15} & \multicolumn{2}{c}{66.67} & \multicolumn{2}{c}{39.80} & \multicolumn{2}{c}{63.47} \\

\bottomrule
\end{tabular}
\end{table*}


\noindent\textbf{Effectiveness of Improving UI Exploration.}
\autoref{tab:sa} shows the improvement brought by \sys's static ad analysis for the \sys and the baseline approaches.
It is easy to observe that \sys's static ad analysis substantially improves the detection rates for all the approaches. 
Specifically, it improves Monkey by $8.86$ percentage points (from $23.99\%$ to $32.84\%$), MadDroid by $7.01$ percentage points (from $26.20\%$ to $33.21\%$), AdGPE by $7.01$ percentage points (from $23.25\%$ to $30.26\%$), and Guardian by $15.50$ percentage points (from $19.19\%$ to $34.69\%$). Notably, \sys exhibits the largest gain, improving by $20.66$ percentage points (from $39.85\%$ to $60.52\%$).
These results demonstrate the effectiveness of \sys's static ad analysis in enhancing UI exploration approaches for ad detection.

The bottom row of \autoref{tab:sa} demonstrates the ad widgets identified by \sys's  static ad analysis. 
Notably, static analysis alone achieves a generally higher detection rate than UI exploration methods, reaching an $81.15\%$ detection rate for AdView widgets.
This is because some ad UIs are not reached during UI exploration, though they are identified through static analysis.   
The relatively higher static detection rate indicates both the feasibility of using static analysis to guide dynamic UI exploration and the potential to further enhance \sys’s overall performance. 


\noindent\textbf{False Positives.}
We also measure the \sys's static ad analysis's precision of detecting ad widgets, which is $52.27\%$. 
This is expected because the inference rules in Subsection~\ref{subsec:adWidgetIdentification} are designed to maximize the detection rate by sacrificing precision.
Besides, these false positives are acceptable because the cost of triggering a falsely reported ad widget in UI exploration is relatively cheap. If no ad appears, we can simply go back.  
These findings demonstrate the advantages of static ad analysis and also reveal the limitations, which can be addressed by \sys's \uinav.



\subsection{RQ3: Ad-Oriented UI Exploration}
\begin{table}[t]
\caption{Ablation study on \sys's \uinav}
\label{tab:ablation}
\centering
\begin{tabular}{@{}lrrrr@{}}
\toprule
\multicolumn{1}{c}{\textbf{Approaches}} & \textbf{Adview} & \textbf{Popup} & \textbf{Native} & \textbf{Micro Avg.}   \\ \midrule
w/o Structural                  & 38.52           & 52.94          & 21.43           & 35.06          \\
w/o Semantic                      & 49.18           & 47.06          & 35.71           & 43.91          \\
w/o WTG                                 & 52.46           & 54.90          & 36.73           & 47.23          \\
w/ KHopWTG                              & 47.54           & 56.86          & 36.73           & 45.39          \\
w/ Gemini-2.5                         & 63.93           & \textbf{60.78} & \textbf{53.06}  & 59.41          \\
w/ Llama-4                            & 60.66           & \textbf{60.78} & 46.94           & 55.72          \\ \midrule

\textbf{\sys}                           & \textbf{68.85}  & 58.82          & 51.02          & \textbf{60.52} \\ \bottomrule
\end{tabular}
\end{table}

In this RQ, we conduct an ablation study to examine the impact of the three types of exploration guidance employed in \sys. The results are presented in \autoref{tab:ablation}. 
The incorporation of the WTG-based and semantic guidance improves \sys's detection rate by $13.28$ percentage points (from $47.23\%$ to $60.52\%$) and $16.61$ percentage points (from $43.91\%$ to $60.52\%$), respectively. Notably, structural guidance yields the largest performance gain of $25.46$ percentage points (from $35.06\%$ to $60.52\%$), demonstrating \sys's effectiveness in empowering the LLM with prior knowledge of ad-specific UI patterns.


Additionally, we replace Dijkstra's algorithm with KHop (we choose $k=2$ as it has the best performance) for WTG representation, denoting it as \textit{KHopWTG}.
KHopWTG informs the LLM about the $k$-hop neighbors of the current UI in the WTG. 
Surprisingly, we observed a decrease in \sys's performance when utilizing KHopWTG. 
Upon manually analyzing the exploration traces, it appears that KHopWTG's information may cause the LLM to navigate in a greedy manner, being stuck in a local neighborhood.  
Instead, Dijiskra's algorithm guides \sys to all reachable UIs with a global view, which cooperates more effectively with the semantic guidance.

Finally, we evaluate the generalizability of \sys across different LLMs by replacing GPT-5 with the state-of-the-art closed-source model (Gemini-2.5 Pro) and open-source model (Llama-4 Maverick). 
Our results show that \sys generalizes well to existing LLMs: Gemini achieves $59.41\%$, similar to GPT-5's $60.52\%$, while Llama-4 is lower than GPT-5 by $4.80$ percentage points.

\subsection{RQ4: In-the-Wild Study}
Conducting a large-scale evaluation of ad regulation violations in the wild has ethical challenges as it will generate invalid impressions and clicks, which constitutes ad fraud and disrupts the advertising ecosystem. 
Fortunately, we were selected for the Specialized Abuse Track of the Meta Bug Bounty Program~\cite{metabugbounty}, which authorizes us to detect abuse on the platform. This allowed us to evaluate the effectiveness of \sys in detecting ad regulation violations in a real-world setting without ethical concerns.

Specifically, we applied AndroGuard~\cite{androguard} to the AndroZoo dataset~\cite{androzoo} to identify apps that integrate the Meta Ads SDK, yielding a dataset of $2,047$ apps. Deploying \sys on this set, we successfully identified $383$ instances of ad regulation violations, including $97$ placement violations, $86$ interaction violations, and $200$ content violations. Given the substantial manual effort involved, we sampled 10 instances per regulation category and verified that each instance indeed violated the reported regulations. Subsequently, we reported the placement and interaction violations to Google Play~\cite{googleAds} and filed a bug report with the Meta Bug Bounty Program  to disclose the content violations. At the time of submission, the report is under review. Due to space limits, we provide the details of this study on our website~\cite{AdSuite}. These results demonstrate \sys's capability to detect actionable regulation violations in the wild.

\section{Discussions}


\noindent\textbf{LLM Unification}.
In this work, we demonstrate the promising effectiveness of LLM in unifying static analysis and dynamic UI exploration. Evaluation results demonstrate \sys's superior performance over both standalone static and dynamic analysis. 
The fundamental idea of our work is not limited to ad widget detection but can be generalized to various mobile GUI testing tasks that target specific app components or APIs, such as mobile payment services.


\noindent\textbf{Static Analysis}. 
In future work, we plan to extend our static analysis to detect these ads by modeling fragments as windows in a WTG by extending \sys with existing static analysis tools~\cite{chen2019storydroid,chen2023deuedroid}.  
Currently, \sys does not support iOS apps as they are closed-source. However, the mobile advertising process depicted in ~\autoref{fig:Mobile Advertising Process} is consistent in both Android and iOS. Therefore, we are able to extend  \sys's framework to iOS by following existing iOS testing work~\cite{wu2023cydios}.

\noindent\textbf{LLM Agents}.
Recent LLM-based agents~\cite{wen2024autodroid,zheng2024gpt} have demonstrated strong reasoning and planning capabilities. 
We believe that integrating such agents could further improve \sys’s ability to detect mobile ads, though it may also introduce additional runtime and resource overhead.
Balancing accuracy and computational cost remains an important direction for future work.




\section{Related Work}

\noindent\textbf{Static Android UI Analysis.} 
Existing tools such as Gator~\cite{Gator,gator2}, Frontmatter~\cite{kuznetsov2021all}, and IconIntent~\cite{xiao2019iconintent} employ the Soot framework~\cite{vallee2010soot} to model the lifecycle of Android applications and analyze their callback functions. This process is used for associating UI widgets with corresponding view classes and event handlers. The scope of such static analysis has been expanded to construct WTGs, as presented by Gator's following work~\cite{gator3}, StoryDroid~\cite{chen2019storydroid}, and Promal~\cite{liu2022promal}, which additionally incorporate analysis of inter-component communications. 
\sys is built upon Gator and IconIntent to effectively analyze widgets and construct WTGs for static ad analysis.

\noindent\textbf{Mobile Ads Detection.}
UI exploration is a major approach for ad content identification~\cite{hu2011automating,rastogi2013appsplayground,nath2013smartads,liu2014decaf,hao2014puma}. 
For example, MadDroid leveraged breadth-first search for UI exploration and rule-based HTTP hooking for ad content identification~\cite{liu2020maddroid}. Researchers also utilized computer vision techniques to detect button edges~\cite{rastogi2016these} or ad-related visual patterns~\cite{cai2023darpa} for identifying ad contents.
However, the UI exploration techniques used by these approaches are for general-purpose testing, causing low effectiveness and efficiency.


\noindent\textbf{Guided UI Exploration.}
Prior research on guided Android UI exploration mainly utilized static analysis~\cite{lai2019goal} and machine learning techniques~\cite{ran2023badge,pan2020reinforcement}. 
More recently, LLMs have been recognized as powerful tools for automating UI exploration. 
Existing work such as Guardian~\cite{ran2024guardian} and AutoDroid~\cite{wen2024autodroid} has demonstrated the efficacy of LLMs in understanding mobile UI and planning exploration paths. 
Inspired by these developments, the design of \sys integrates a hybrid approach that synergizes both LLM capabilities and static analysis. This combination aims to enhance the effectiveness of guided UI exploration.

\noindent\textbf{Mobile Ads Analysis.} 
Research on mobile ads analysis spans various directions.
A line of studies investigated the performance and extra energy consumption of mobile ads~\cite{vallina2012breaking,khan2013cameo,Mohan2013}, while others proposed developer-side solutions to enhance mobile ad quality~\cite{Gui2015, jin2019madlens}.
Another key direction of research addressed privacy and security concerns, including privacy leakage~\cite{grace2012unsafe, enck2014taintdroid, hardt2012privacy, liu2015efficient}, mobile ad fraud~\cite{liu2014decaf, dave2013viceroi}, and malicious advertising~\cite{rastogi2016these, liu2020maddroid}. By detecting ad widgets, \sys supports various downstream analyses in these directions.
\section{Conclusion}
In this paper, we have proposed a novel approach for mobile ad regulation named \sys that innovatively harnesses LLMs to unify static ad analysis and dynamic UI exploration with ad widget placement strategies and knowledge of ad UIs. 
Our evaluation on $271$ ad widgets demonstrates the superior performance of \sys, improving the detection rate by $25.83$ percentage points over the strongest baseline. From the detected ads, \sys further identifies $133$ instances of regulation violations, exceeding state-of-the-art approaches by $34.34\%$, highlighting its effectiveness in assisting ad regulation.

\bibliographystyle{IEEEtran}
\bibliography{bibfile}

\end{document}